# Trionic Optical Potential for Electrons in Semiconductors

Martin J. A. Schuetz, Michael G. Moore & Carlo Piermarocchi

*Department of Physics and Astronomy, Michigan State University, East Lansing, Michigan, 48824*

**Laser-induced optical potentials for atoms have led to remarkable advances in precision measurement, quantum information, and towards addressing fundamental questions in condensed matter physics. Here, we describe analogous optical potentials for electrons in quantum wells and wires that can be generated by optically driving the transition between a single electron and a three-body electron-exciton bound state, known as a trion. The existence of a bound trion state adds a term to the ac Stark shift of the material proportional to the light intensity at the position of the electron. According to our theoretical calculations, this shift can be large relative to the thermal equilibrium temperature of the electron, resulting in a relatively strong optical potential that could be used to trap, guide, and manipulate individual electrons within a semiconductor quantum well or wire. These potentials can be thought of as artificial nano-structures on the scale of 100 nm that can be spin-dependent and reconfigurable in real-time. Our results suggest the possibility of integrating ultrafast optics and gate voltages in new resolved-carrier semiconductor opto-electronic devices, with potential applications in fields such as nano-electronics, spintronics, and quantum information processing.**

The optical properties of one- and two-dimensional semiconductor confined systems are dominated at low temperature by bound electron-hole pairs known as excitons [1]. In samples with a small excess of electrons, an exciton can capture an additional electron to form a charged exciton, or trion, which is the bound state of two electrons and a hole [2,3]. In good III-V and II-VI quantum well (QW) samples, trions with high mobility have been observed [4-6]. From an atomic physics perspective, the electron to trion optical transition occurring in a semiconductor host environment effectively gives a second internal state to the electron, which is otherwise a point-like, structure-less particle in free space. Optically driving the transition between these two states then induces the electron to respond to the light an atom-like manner.

When transitions between two internal atomic energy levels are driven by near-resonant laser light, the ground-state ac Stark shift is proportional to the laser-intensity at the position of the atom. Thus a spatially varying intensity pattern creates a spatially varying optical potential, which acts on the atomic center-of-mass motion. We propose creating analogous optical potentials for semiconductor charge carriers confined to one- or two- dimensions by driving the sample with light tuned near the trion resonance. A spatially inhomogeneous intensity profile, e.g. a standing wave interference pattern, will result in a mechanical potential acting on the electron (see Fig. 1). As a resonant effect, the trionic optical potential is many orders of magnitude stronger than the ponderomotive potential [7] that arises from the Lorentz force acting on the electron. In contrast to electron-hole pairs localized by transversal light forces [8], an electron (or hole) in a trionic optical potential will not radiatively decay, opening the possibility for optical manipulation of charge flow and/or the creation of interesting many-electron states, i.e. a solid-state quantum simulator [9].

At present, we focus on the case of electrons confined GaAs and CdTe QWs, whose optical properties have been most thoroughly investigated experimentally, but similar potentials should be obtainable for holes, as well as for carriers confined in quantum wires. A variational approach has been used to calculate the wave-function and the optical properties of the three-particle trion state, as described in Methods. Under laser excitation, the electron makes virtual transitions into a bound trion state; while at the same time, the laser drives the transition to non-localized excitons, whose wave functions are modified by the electron-exciton interaction; hence they can be viewed as 'unbound trions'. Including both types of transitions we have treated the light matter coupling perturbatively and obtained a second-order effective Hamiltonian. We consider a monochromatic standing wave $E(r,t) = E_0 \cos(Qr)\cos(\omega t)$ containing photon modes with $\pm Q$ in-plane momenta, producing an intensity pattern in the form of periodic stripes along the QW. Neglecting small non-local effects (see Methods and Fig S1), the resulting potential for the electron can be expressed as

$$U(r) = -\varepsilon \Delta_t f_c(\Delta_t) \cos^2(Q \cdot r), \tag{1}$$

where we introduced the dimensionless saturation parameter $\varepsilon = \chi (\Omega_0 / \Delta_t)^2$. The parameter $\Omega_0 = d_0 |E_0|$ is the Rabi energy, with $d_0$ being the inter-band dipole moment, and $\Delta_t$ is the difference between the trion resonance and the laser energy. The factor $f_c(\Delta_t)$ is due to the presence of the continuum of unbound trions (see Methods), and can be approximated by $f_c(\Delta_t) = \dfrac{E_b}{E_b + \Delta_t}$, where $E_b$ is the trion binding energy. Since unbound excitons in our system would be equivalent to vacuum excitations in the atomic case, we interpret this coefficient as a vacuum polarizability correction. In the definition of the saturation parameter we include the coefficient $\chi$ (see Fig. S2), which is proportional to the integral over all configurations of the three particle wavefunction with one electron and the hole taken at the same position [10]. The coefficient $\chi$ is about 80 for GaAs and 90 for CdTe.

By decreasing the saturation $\varepsilon$, heating of the electron due to spontaneous scattering of photons can be reduced. Photon scattering is enhanced by the trion resonance, occurring at the rate $\Gamma_{se} = \varepsilon \Gamma_t$, with $\Gamma_t \sim 10^{10} \, s^{-1}$ being the natural line-width of the trion state. While $\Gamma_{se}$ scales as $\sim 1/\Delta_t^2$, the trap depth scales as $f_c(\Delta_t)/\Delta_t$. The sign of the trion-photon detuning $\Delta_t$, corresponds to two qualitatively different regimes for the optical potential (see Fig. 2). For red detuning, $\Delta_t > 0$, the potential minima are at the anti-nodes of the intensity profile, so that the electrons will be attracted towards high field intensities, where $\Gamma_{se}$, proportional the laser intensity, is maximum. The situation is reversed for intermediate blue detuning, $\Delta_t < 0$, where the potential minima correspond to the nodes of the standing wave pattern, where $\Gamma_{se}$ is minimized. We show in Fig.3 (a) and (b) the potential depth $U_0$ for red and blue detuning as a function of the controllable parameter $\varepsilon$. The detuning $\Delta_t$ as well as the potential depth is represented in units of the trion binding energy $E_b$, which is typically of the order of 2 meV in GaAs and 3.6 meV in CdTe semiconductor QWs. For red detuning, the continuum and the bound trion level give contributions of opposite sign to the shift of the electron ground state energy (see Fig. 2). As a consequence, the potential depth saturates at a maximum of $\varepsilon E_b$. For intermediate blue detuning, $E_b < |\Delta_t| < 0$, the continuum and bound trion contributions add constructively, resulting in a significantly deeper potential, scaling as $1/(E_b - |\Delta_t|)$. The enhancement due to the singularity at $\Delta_t = -E_b$ is limited by the requirement to remain well-below the exciton resonance, whose natural line-width, $\Gamma_X$, for both GaAs and CdTe is approximately $0.05 E_b$. For $\varepsilon = 0.05$, we find that a potential depth of $0.5 E_b$ is feasible. To realize this strong potential, laser intensities of $I = 3 \cdot 10^3$ W/cm$^2$ and $I = 9 \cdot 10^3$ W/cm$^2$ would be required for GaAs and CdTe QW, respectively. These intensities are much weaker, for instance, than the ones used to observe the exciton Stark shift [11], where a strong laser was followed by a weak probe to probe the energy *difference* between adjacent light-dressed states [12]. The optical potential, on the other hand, is simply the shift of the lowest dressed state, and is therefore much stronger.

A natural energy scale for quantifying the effects of laser heating is the single photon recoil energy defined as $E_R = \hbar^2 Q^2 / 2m_e^*$, i.e. the kinetic energy of an electron with a momentum equal to that of the laser photons and mass equal to the electron effective mass $m_e^*$. Translational invariance along the QW implies conservation of the in-plane momentum only, which reduces the average recoil kick an electron experiences in the process of spontaneous photon emission. For GaAs and CdTe the in-plane kinetic energy imparted by photon recoil, averaged over the emission angle, are $\langle E_R \rangle = 0.29$ meV and $\langle E_R \rangle = 0.16$ meV, respectively. The small effective mass of the electron - which is about seven orders of magnitude smaller than the mass of rubidium atoms - gives rise to relatively large recoil energies. However, this energy has to be compared to the maximum achievable depth of the trionic optical potential, which is significantly deeper than atomic potentials due to a much higher oscillator strength, and the enhancement due to the background polarizability. We compare our potential-depth to recoil-energy ratio with three typical atomic optical lattice experiments on macroscopic quantum interference [13], Bloch oscillations [14], and quantum phase transitions [15]. The maximum achievable potential depths in these experiments have been $2.1 E_R$, $6 E_R$ and $22 E_R$, respectively. Despite the large recoil energies involved in our system, we can still predict potential depths of about $3 E_R$ and $10 E_R$ for GaAs and CdTe, respectively. For CdTe, we have also estimated the number of quasi-bound states by approximating the trapping potential as a single harmonic oscillator and comparing the energy level separation to the total depth. We found that three levels are bound within a large range of detuning and saturation parameters.

In contrast to the case with trapped atoms, cooling of the electron is provided by a mechanism completely separate from the optical trapping mechanism: coupling to the phonon reservoir. In our energy range, the electrons will cool down by acoustic phonon emission (Fig. S3). The equilibrium temperature is therefore determined by competition between phonon emission and laser heating (we assume that the experiment will be carried out at 300 mK, where phonon absorption can be neglected). The laser heating rate $R_h$ - measured in energy per time - can be expressed as $R_h = \varepsilon \Gamma_t \langle E_R \rangle$, while acoustic phonon emission tends to cool the electron down at some cooling rate $R_c(E)$.

The two competing processes lead to an effective equilibrium energy defined by $R_c(E_{eq}) = R_h$ (Fig. S4). Using the electron-phonon deformation potential interaction, a lattice temperature of 300 mK and a QW width of 20 nm we have calculated cooling and heating rates and the effective equilibrium electron temperature for different values of the saturation $\varepsilon$ (Table 1). For $\varepsilon = 0.05$ and $\Delta_t = -0.9 E_b$, we predict a potential depth of $0.5 E_b$, corresponding to 1 meV for GaAs and 1.8 meV for CdTe. The depth of the trapping potential thus exceeds the electron equilibrium temperature by up to one order of magnitude.

In the blue-detuned case, the possibility arises for phonon-assisted photon absorption. Virtual trions created by photon absorption can decay into real trions by acoustic phonon emission. This occurs at a rate of the order of .8 ns$^{-1}$ in homogeneously illuminated GaAs samples with $\varepsilon = 0.05$ (Fig. S3). We note that, in the blue-detuned case electrons are confined in the vicinity of the nodes of the intensity pattern, which will further suppress both the photon and phonon emission rates by a factor that can be analytically calculated as $\left(1 - e^{-\sqrt{E_R/U_0}}\right)/2$. Note that while the resulting real trion will decay by spontaneous photon emission, and thus lead to an increase in the heating rate, the emitted photons will be at the trion resonance frequency. This raises the possibility of measuring the electron density by imaging the light at the trion resonance frequency, as this light will only be emitted in proximity to an electron.

As a result of spin-orbit interaction and confinement, the trion level structure allows spin-up electrons to respond only to right-handed photons, while spin-down electrons see only left-handed photons (see Fig. S5). This means that the trionic optical potentials for spin-up and spin-down electrons, generated by left- and right-circularly polarized laser fields, respectively, can be completely independent of each other. Thus it should be possible to generate entanglement between an electron's spin and orbital motions. This could then be used to generate entanglement between the spins of two neighbouring electrons, whose separation, and therefore interaction strength, can depend strongly on the spin-states of the individual electrons. These possibilities show that, when viewed either as a platform for addressing fundamental open questions in condensed matter physics, and/or as a platform for spin-based quantum information processing, optically trapped charged particles in a semiconductor environment are not intrinsically disadvantageous with respect to their atomic cousins.

In future work, we will consider optical potentials for heavy holes, which have a reduced heating rate due to their larger mass; as well as for bi-layer excitons [16] which have the advantage of being electrically neutral, and thus interact at a much shorter range than either electrons or holes. Also, we will explore the possibility for even deeper potentials in quantum wires, where confinement in two dimensions should lead to even stronger enhancement.

**Methods**

**Charge loading:** The loading of charges in the optical potential could be experimentally realized using different techniques. The electronic density has to remain low in order to limit Coulomb effects, ideally around $10^9$ cm$^{-2}$, corresponding to a Fermi energy of about 0.02 meV, much smaller than the optical potential depth. The quantum well could be embedded in a Schottky diode structure controlled by external dc field. This method has been used in loading layers of quantum dots with one electron each [17]. Realizations relying on delta doping layers may present problems due to ionized doping centers, which produce an additional random potential on the quantum well. However, according to direct scanning probe techniques, this potential is characterized by patterns with dominant length scales of the order of one micron [18] and we expect that electron trapping could be created inside these micron-sized patterns. Since we need a low carrier density, the sample design can be optimized to minimize the effect of ionized impurities. For instance, the dopant layers could be made thick and relatively distant from the optical potential region. Charge loading could also be realized by photo-doping double quantum wells systems. Here, spatially separated electrons and holes can be created following high energy excitation. For distant quantum wells the charges can remain confined for seconds [19] and we could obtain two independent optical potentials for the two types of carriers. For smaller separation of the wells the optical potential will trap indirect excitons and could complement other methods of controlling indirect exciton fluxes in integrated circuits [16]. Finally, one could simply take advantage of unintentional doping in the barriers, and pump with an energy just above the ionized acceptor to conduction transition in the barriers to produce electrons that relax in a single quantum well, as recently demonstrated [20].

**Competing effects:** We have studied the possibility of optical excitation of the electron out of the confining quantum well. This process competes with virtual trion formation, but the corresponding matrix element is vanishingly small due to the mismatch between the two dimensional wavefunction in the quantum well and the extended bulk-like states in the barriers. An upper bound estimation of the branching ratio between electron ionization and trion creation in the QW, neglecting Coulomb enhancement effects, gives $10^{-5}$. We have also considered disorder effects due to fluctuations in the number of atomic mono-layers. These fluctuations create islands corresponding to different well or wire widths. The optical potential will not be affected if the characteristic length of the islands is much smaller or much larger than the carrier confinement length. From this point of view, a promising system for the experimental realization is represented by V-groove semiconductor quantum wires, in which 1D islands of the order of 1 micron and extremely narrow homogeneous emission lines have been explicitly observed [21,22]. In wires the trion binding energy and thus the optical potential will be larger than the one calculated above. In principle very high-quality quantum wells without growth interruption should feature large islands similar to the ones in quantum wires. For relatively large quantum wells with width around 20 nm, the exciton linewidths can be extremely narrow. For instance, 25nm quantum wells with exciton lines of 0.075meV have been reported [23]. In those samples the effect of interface disorder is completely negligible. Alloy disorder is also present, but due to its extremely short range characteristic length [24] its effects are averaged out within the electron envelope function. Finally, unintended doping can contribute to the absorption through impurity-bound exciton resonances and their phonon satellites lines. However, absorption of energy due to unintended doping mostly occurs in the barriers and substrate, and is significant only at energies quite higher than the quantum well trion resonance.

**Detection schemes:** The detection of electrons trapped by the optical potential can be realized by detecting the weak optical emission at the trion energy which will be spectrally separated than the pump energy. The emission at the trion energy at a given spot is a signature of the presence of the electron. In the blue detuning case the trion emission will occur at the dark spots of the pump interference pattern and should be less challenging to observe. Both far field and near field optical techniques should be able to demonstrate the carrier localization. In contrast to atomic systems, the confined carriers are charged, and charge imaging methods could also be used to detect electrons trapped by the optical potential in a way similar to the detection of single electrons trapped by impurity centers [25]. We note that the experiments we propose are very similar to Four Wave Mixing experiments that have been carried out in 2DEG systems (for a review see [26]). The only difference is that we are proposing to explore a different regime of carrier density and excitation energy.

**Variational Calculation:** For the trion state (X⁻) we have used a two dimensional variational Hylleraas function [27] of the form $\varphi_b(s,t,u) = N e^{-\alpha s}(1+\beta u + \gamma t^2)$, where $\alpha$, $\beta$ and $\gamma$ are three variational parameters and $N$ is the normalization constant. The position of the three carriers in the trion is expressed in terms of two elliptic coordinates $s = r_{e_1 h} + r_{e_2 h}$ and $t = r_{e_1 h} - r_{e_2 h}$ and the inter-electron distance $u = r_{e_1 e_2}$. Hylleraas-type wave functions are known to describe well radial and internal angular correlation effects [28]. For GaAs and CdTe quantum wells we have obtained trion binding energies of about 2.05 meV and 3.6 meV respectively, which is in good agreement with the experiments [6]. Using this wavefunction we have found radiative lifetimes of about 20 ps for both materials, also in a reasonable agreement with the experimental values [5].

The matrix element for the optical transition from an initial electron with momentum $k$ to a bound trion state with centre-of-mass momentum $K$ can be written as

$$\langle K | H_{LM} | k \rangle = \frac{\Omega_0}{2}\left[\delta_{K,k+Q} I(\beta_X k - \beta_e Q) + \delta_{K,k-Q} I(\beta_X k + \beta_e Q)\right], \qquad (2)$$

where the first and second terms on the right hand side describe the absorption of a photon in the $+Q$ mode and in the $-Q$ mode, respectively. The coefficient $\beta_e = 1-\beta_X = m_e/m_t$ is the electron to trion mass ratio, the light matter interaction is indicated by $H_{LM}$ and $\Omega_0$ is the Rabi energy. The quantity $I(p) = \frac{1}{\sqrt{2\pi}} \int dr\, e^{-ipr} \varphi_b(r,r,r)$ is the Fourier transform of the bound trion wave function $\varphi_b$ taken with one electron and the hole at the same position. This function enters in Eq. (2) with $p = \beta_X k \pm \beta_e Q$, which is the relative momentum of the initial electron $k$ and the photo-created exciton with $\pm Q$. For the calculation of the optical potentials we have used a full analytical expression for the function $I(p)$ [29]. Note that the enhancement factor is approximately given by $\chi = |I(p=0)|^2$.

**Perturbation theory:** The second order effective Hamiltonian for the electron can be written as

$$H_{eff} = \sum_k \left(\varepsilon_k^{(0)} + \varepsilon_k^{(2)}\right)\left|k^{(0)}\right\rangle\left\langle k^{(0)}\right| + \varepsilon_k^{(0)}\left(\left|k^{(0)}\right\rangle\left\langle k^{(2)}\right| + \left|k^{(2)}\right\rangle\left\langle k^{(0)}\right|\right), \quad (3)$$

where the subscripts (0) and (2) indicate the zeroth and second order corrections to the electron energy $\varepsilon_k$ and wave-function $|k\rangle$ with respect to the light matter coupling. Terms involving the first order corrections, whose eventual projection on electron position eigenstates $|r\rangle$ give zero because of orthogonality, can be neglected. The second term on the right hand side of Eq. (3) includes off-diagonal processes between electronic states with different $k$ due to the mixing with photon modes. These off-diagonal contributions can be written in the form

$$H_{OD} = \left[-\sum_k \frac{\Omega_0^2}{4}\left(\frac{1}{\Delta_t} + \frac{\chi_c(\Delta_x)}{\Delta_x}\right)|I(k)|^2 + \frac{\Omega_0 \Omega_X}{\Delta_x} O(K,k)I(k)\right]|k-Q\rangle\langle k+Q| + hc \quad (4)$$

where $\Delta_{x(t)}$ is the photon exciton (trion) detuning, $O(K,k) = \langle K | K, k \rangle$ is the overlap between a bound state trion with centre of mass momentum $K$ and an unbound state composed of one exciton with centre of mass momentum $K$ and one electron of momentum $k$ and $\Omega_X$ is the Rabi energy modified by excitonic effects. We have defined the quantity $\frac{\chi_c(\Delta)}{\Delta} = \sum_k \frac{O(0,k)^2}{\hbar^2 k^2 / 2\mu_t + \Delta}$, where $\mu_t$ is the reduced mass of the free exciton-free electron system. These off-diagonal terms are the dominant terms in the derivation of a potential which carries the signature of the intensity profile. From the effective Hamiltonian in $k$-space, we derive a Schrödinger equation for the electron wavefunction $\psi(r)$ in the presence of the standing wave given by

$$-\frac{\hbar^2 \nabla^2}{2m_e^*}\psi(r) - \frac{\Omega_0^2}{\Delta_t} f_c(\Delta_t) \int dr' \cos^2\left(Q\frac{r+r'}{2}\right) m(r-r')\psi(r') = E\psi(r). \quad (5)$$

We note that the optical potential is not diagonal in coordinate representation; rather it depends on the nonlocal kernel $m(r - r')$. The continuum factor and non-local kernel are new effects related to the vacuum polarizability of the electron's environment, and are not observed in atomic systems. A plot of the kernel function $m(x)$ is shown in Fig.

S1, where we see that it matches the size of the trion bound state. As the size of the kernel is small compared to the optical wavelength, nonlocal effects can be treated as a perturbation that effectively averages the light intensity over a small area centered on the electron position, corresponding to the approximation $m(\boldsymbol{x}) = \chi \delta(\boldsymbol{x})$.

We thank Prof. C.W. Lai and S. H. Tessmer for discussions on the experimental realizations. We acknowledge support by the National Science Foundation (C.P., M.M., and M. S.), the German Studienstiftung Program (M.S.), and the Fulbright Foundation (M.S.). Authors' Contributions: M.S. performed calculations; C. P. and M. M. proposed concept; C. P., M.M., and M. S. wrote the manuscript.

Correspondence and requests for materials should be addressed to M.S. (schuetzm@msu.edu).

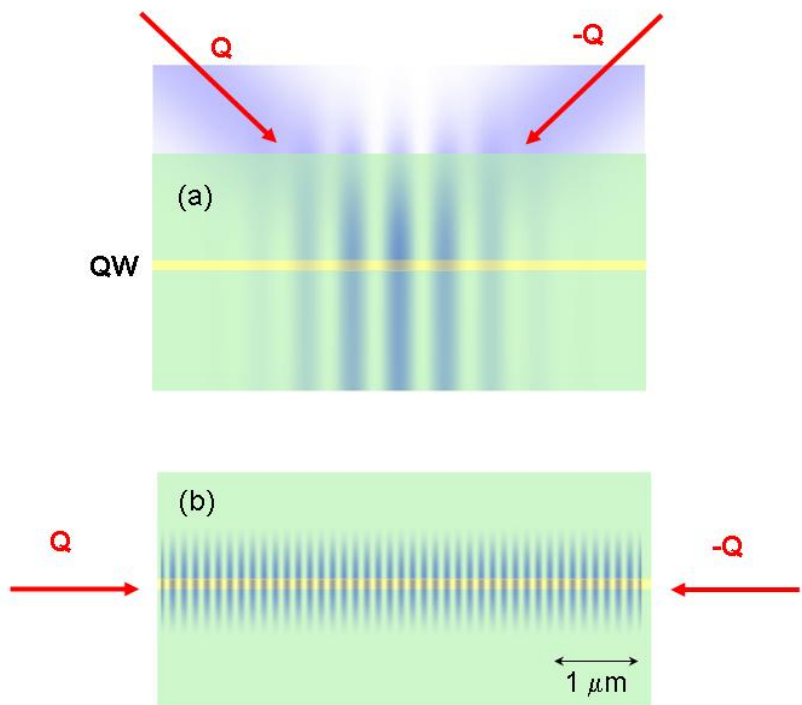

FIGURE 1. Interference pattern for two laser beams incident on the semiconductor structure. (a) Top excitation at incidence angle $\pi/3$. (b) Lateral excitation. The interference pattern has been calculated using an average index of refraction n=3.5 corresponding to typical III-V based structures. Note that the optical lattice periodicity can be partially controlled by changing the incidence angle.

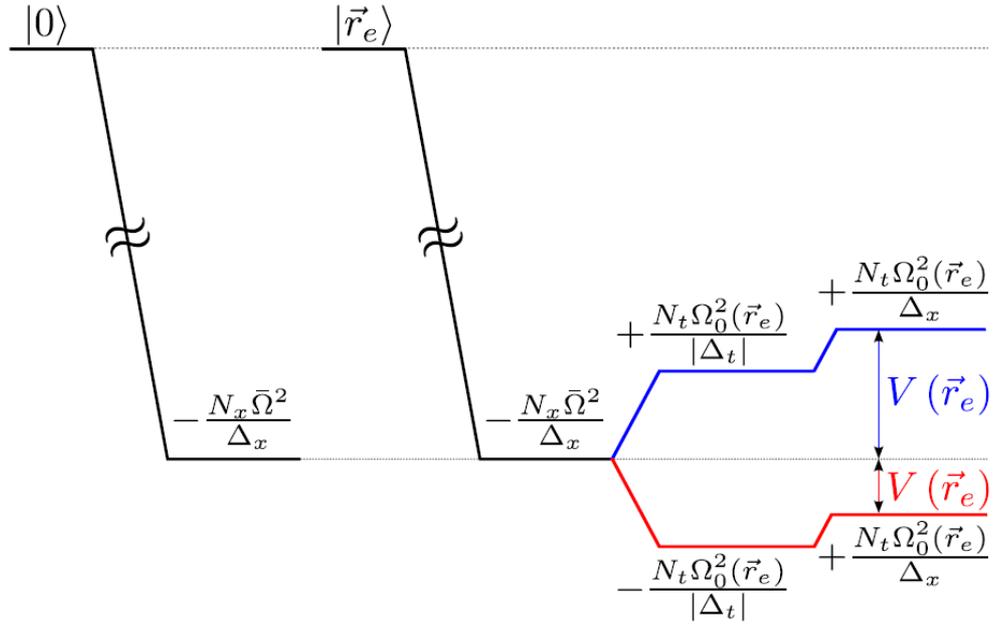

FIGURE 2. Scheme of energy level shifts for an empty quantum well $|0\rangle$ and for a quantum well with one electron at $\vec{r}_e$, in the red and blue detuning case. An exciton in a neutral QW, ideally coherent over the whole sample, has an oscillator strength that scales as $N_x = A/A_x$, where $A$ is the total area of the sample and $A_x$ is the exciton size. In contrast, the fact that a free carrier must be already present in the sample in order to create a trion implies that the trion oscillator strength scales as $A_t/A_x = N_t$ [10], where $A_t$ is the typical trion size. In the diagram the contributions from the bound trion (proportional to $\Omega_0^2/\Delta_t$), and the continuum (proportional to $\Omega_0^2/\Delta_x$) are shown separately. $\Delta_{x(t)}$ is the photon exciton (trion) detuning. Notice that the constant global energy shift $-N_x\bar{\Omega}^2/\Delta_x$, where $\bar{\Omega}$ is a spatially averaged Rabi energy does not contribute to the electron potential.

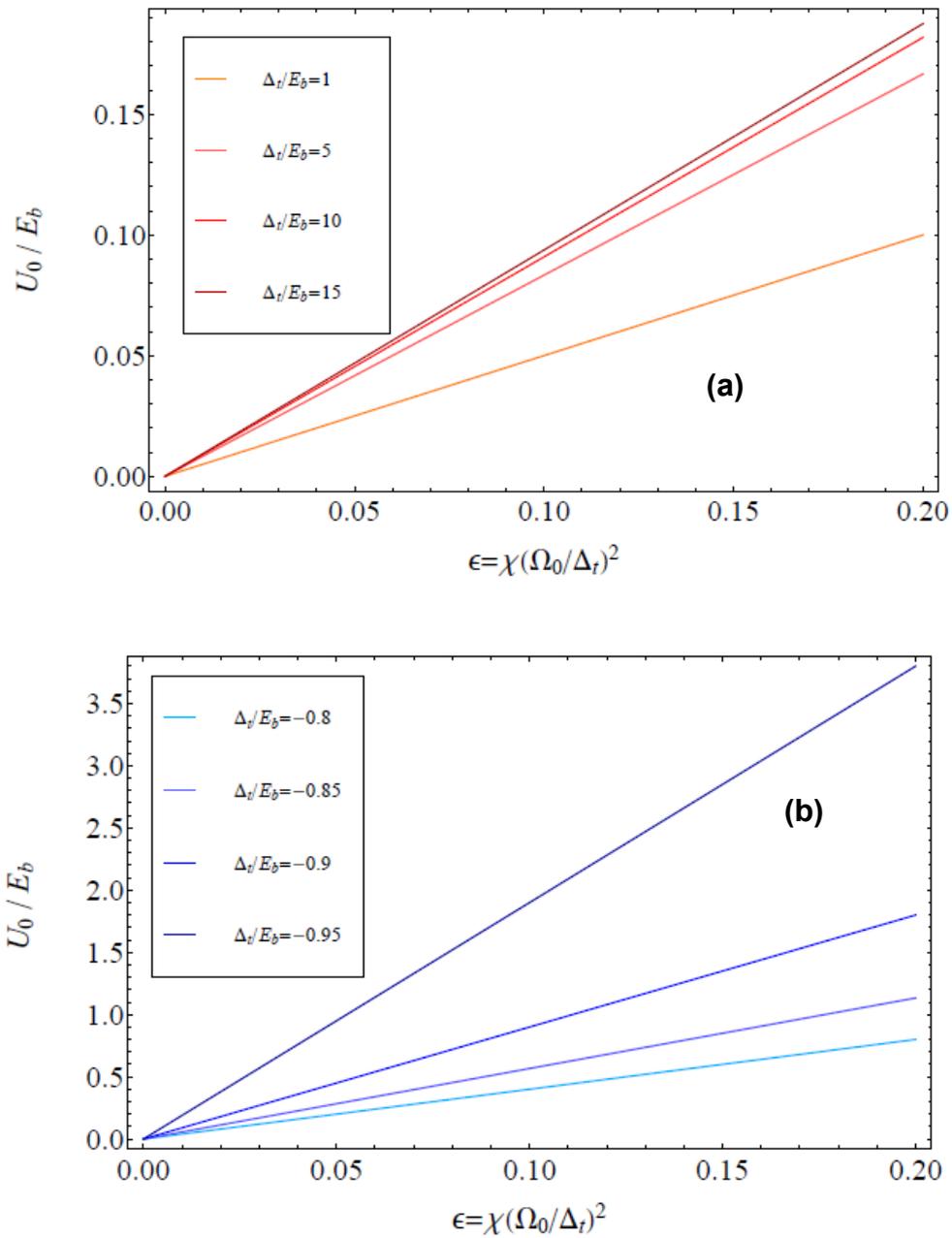

FIGURE 3 Optical potential depths for red (a) and blue (b) detuning expressed in terms of the trion binding energy for different values of the optical detuning with respect to the trion resonance. In the red detuned case the potential is attractive and electrons will be trapped at the anti-nodes of the interference pattern, while for blue detuning the potential is repulsive and electrons will be trapped at the nodes.

**Table 1. Electron equilibrium temperature in K for GaAs and CdTe for different values of the saturation parameter.**

|  | GaAs | CdTe |
|---|---|---|
| $\varepsilon = 0.10$ | 4.6 K | 2.9K |
| $\varepsilon = 0.05$ | 3.5K | 2.3K |
| $\varepsilon = 0.01$ | 2.3K | 1.2K |

TABLE 1. Equilibrium temperature resulting from the balance of laser heating and phonon cooling for different saturation parameters for the two materials. The lattice temperature is 300 mK (see also Fig. S4).

# Supplemental Figures

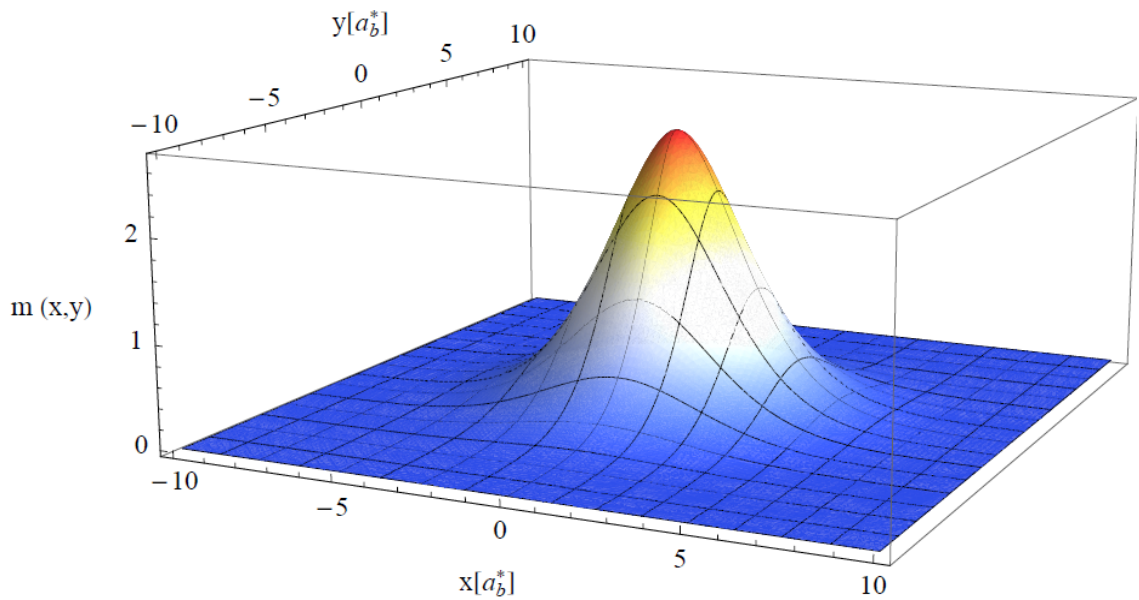

FIGURE S1

FIGURE S1 **Non-locality of the optical potential.** Plot of the nonlocal kernel $m(x, y)$ in the optical potential. The kernel has the same spatial extension of the trion wavefunction and is about $3\, a_b^*$, where $a_b^*$ is the donor Bohr radius $a_b^* = \varepsilon_0 \hbar^2 / m_e^* e^2$, $m_e^*$ is the electron mass, and $\varepsilon_0$ the static dielectric constant.

The non-locality in the electron optical potential can be understood by noting that during an electron-trion virtual transition, the initial position of the electron does not correspond to the center-of-mass position of the resulting trion. The non-locality of the potential thus reflects the effect of inter-particle forces displacing the electron during the virtual excitation. In principle, optical potentials for atoms also have non-local features, e.g. arising from the momentum differential between the excited and ground states due to photon recoil. However, due to the large atomic mass and the small atomic size relative to the optical wavelength, such effects are too small to be observed.

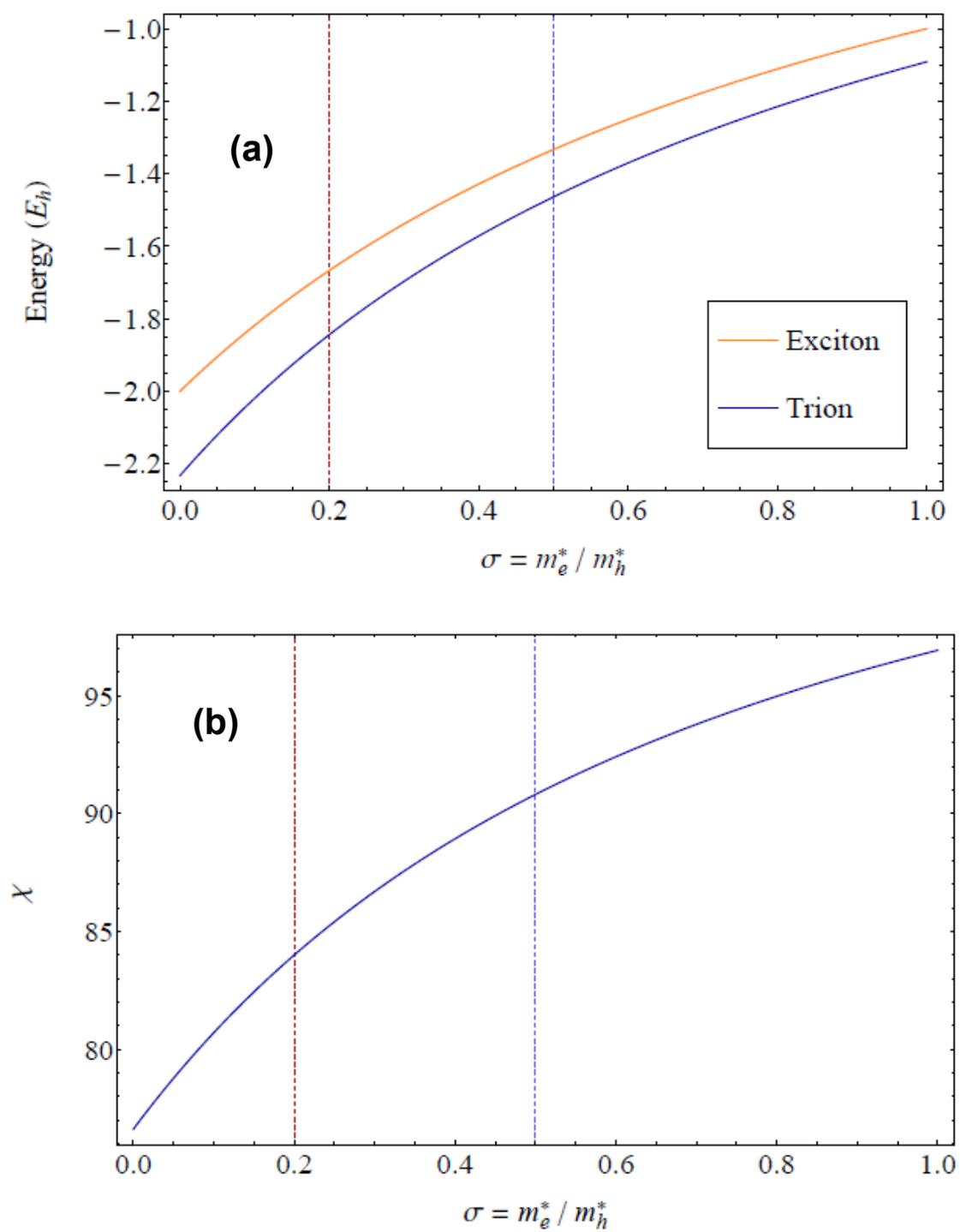

FIGURE S2

FIGURE S2 **(a) Variational Calculation: Exciton and Trion energy.** Exciton binding energy (orange) and Trion formation energy (blue) as a function of the mass ratio $\sigma = m_e^*/m_h^*$ in units of donor Hartrees $E_h = \frac{m_e^* e^4}{\varepsilon_0 \hbar^2}$, where $m_h^*$ is the hole mass. The trion formation energy is the trion energy relative to the total energy of two free electrons and one free hole. The dashed vertical lines indicate the mass ratio values for GaAs (red) and CdTe (blue). The trion is stable against dissociation into a free carrier and one exciton for every mass ratio $\sigma$. The variational wavefunction has the form $\varphi_b(s,t,u) = N e^{-\alpha s}(1 + \beta u + \gamma t^2)$ (see Methods), where $\alpha$, $\beta$ and $\gamma$ are variational parameters and $N$ is the normalization constant. For GaAs with $\sigma = 0.2$, and CdTe with $\sigma = 0.5$ we have obtained the values $\alpha = 1.293$, $\beta = 0.396$, $\gamma = 0.374$, and $\alpha = 1.008$, $\beta = 0.260$, $\gamma = 0.233$, respectively. Coordinates in the trial wavefunction are expressed in units of donor Bohr radius $a_b^* = \varepsilon_0 \hbar^2 / m_e^* e^2$, which defines a natural length scale for the problem. Typically, $a_b^*$ is 5-10 nm. The trion binding energy $E_b$ is the difference between the orange and the blue curve. Note that when converted to conventional units, CdTe has a larger binding energy since $E_h$ in CdTe is more than two times bigger than in GaAs. The trion binding energies from the variational calculation are 2.05 meV and 3.6 meV for GaAs and CdTe, respectively, in a very good agreement with previous theoretical studies [1-3] and experiments [4-7]. **(b). Trion enhancement factor $\chi$.** In the electron-to-trion transition the bare semiconductor inter-band dipole moment $d_0$ is enhanced by a factor $\sqrt{\chi}$. This takes into account that a bound trion forms if the additional electron-hole pair is created within a finite range of the excess carrier. This range is determined by the relative trion wavefunction, and $\chi$ is proportional to the integral of the wavefunction over all three particle configurations that will have one electron and the hole at the same position [8]. Partly owing to $\chi$, the oscillator strength for the electron-trion transition,

proportional to $\chi d_0^2$, exceeds its atomic counterpart by three to four order of magnitude since $d_{trion}^2 = \chi d_0^2 \approx 10^3 - 10^4 d_{atom}^2$.

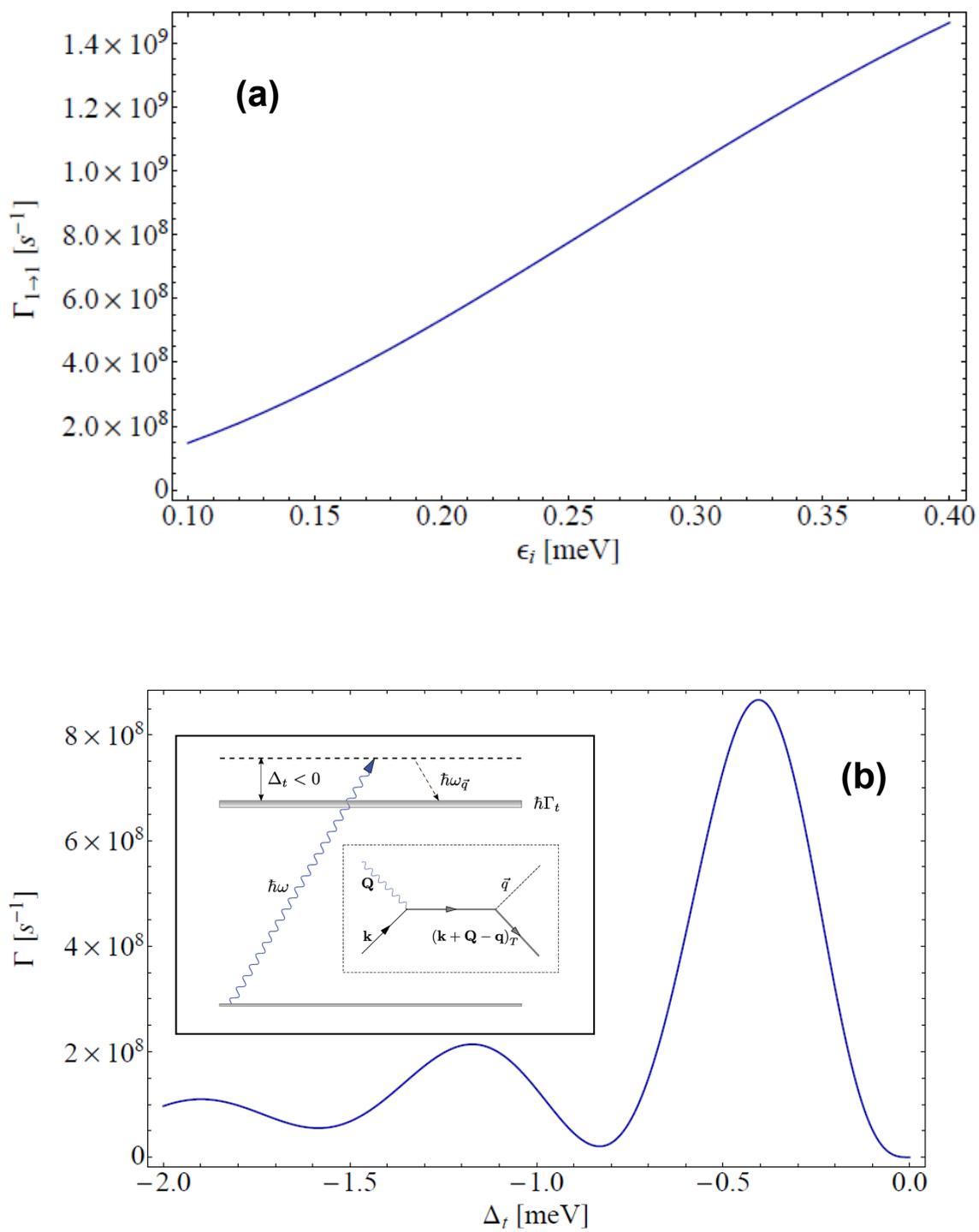

FIGURE S3

FIGURE S3 **(a) Phonon effects: Intraband phonon cooling rate.** Electrons that have been excited by laser heating or Auger processes will relax to the thermal equilibrium energy by emission of acoustic phonons. We show here the calculated intra-band relaxation rate $\Gamma_{1\to 1}$ as a function of the initial electron energy for a 20 nm CdTe quantum well. Only intra-subband contributions have been considered and the calculations were done using electron acoustic phonon deformation potentials and a parabolic band. Note that we are assuming that the experiments will be carried out at temperatures of the order of 300mK, for which absorption processes can be neglected. **(b) Phonon-assisted trion creation.** Trions can be created by a second order process that involves the virtual excitation of a trion by photon absorption and the subsequent absorption or emission of an acoustic phonon. A schematic picture of the process with the relative diagram is show in the inset. Phonon-assisted trion formation could in principle affect the optical potential in the blue detuning case. In the red detuning case, only phonon absorption can contribute to this process, which is suppressed at low temperature. The figure shows the rate for this process calculated for a homogenously illuminated GaAs 20 nm QW with $\varepsilon = \chi |\Omega_0 / \Delta_t|^2 = 0.05$. With $\varepsilon$ held constant the rate first increases as a function of the detuning due to the energy dependence of the carrier-phonon coupling. After reaching a maximum, the rate decreases since for larger detuning values, the in-plane phonon momentum necessary for energy-momentum conservation exceeds the inverse of the characteristic trion size. Note that even in this homogeneous case this rate is much smaller than the radiative recombination rate, which is about 5 $10^{10}$ Hz, so the population of real trions in the system will remain small in the presence of a continuous laser excitation. Furthermore, in the case of a blue detuned optical potential, which is repulsive, electrons are attracted towards the dark spots of the intensity pattern so the effect of phonon-

assisted trion creation on the optical potential seen by the electron will be reduced. The reduction factor can be analytically calculated as $\left(1-e^{-\sqrt{E_R/U_0}}\right)/2$ where $E_R$ is the recoil energy and $U_0$ is the potential depth.

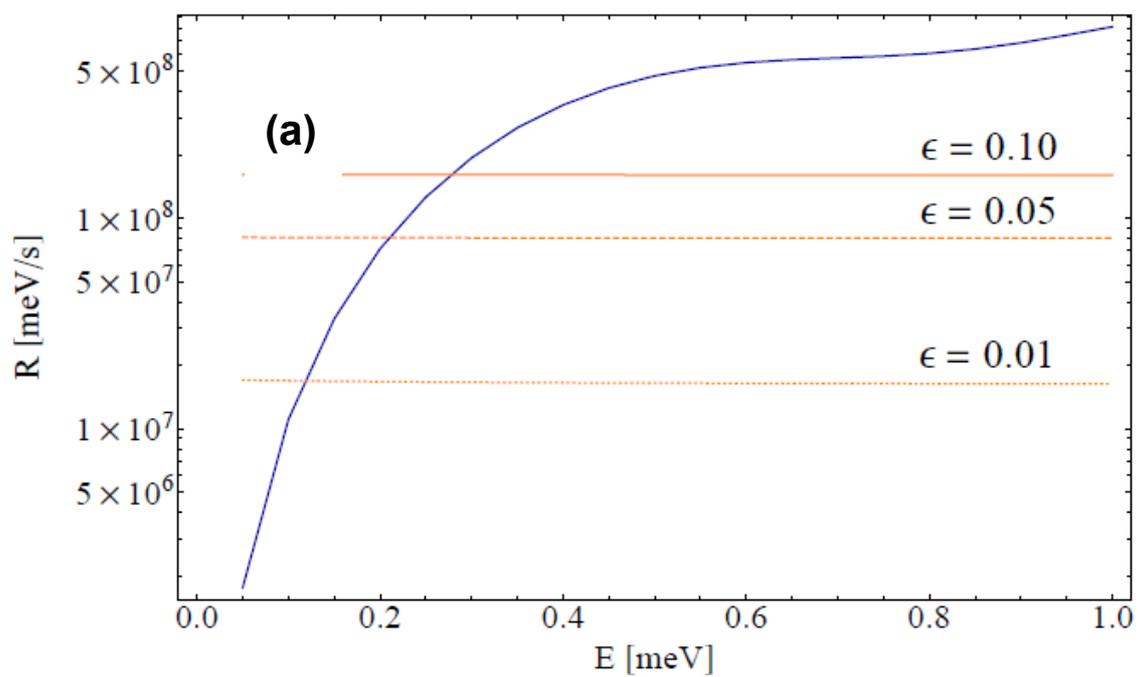

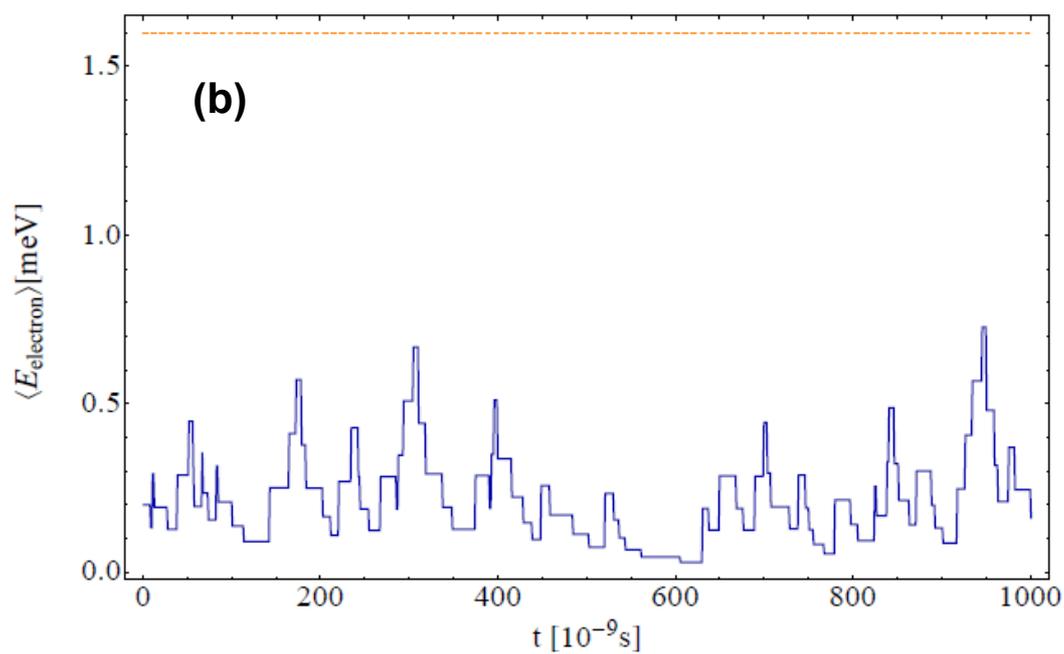

FIGURE S4

FIGURE S4 **(a) Phonon cooling and Photon heating: Equilibrium temperature**. An electron trapped in an optical potential will experience two competing effects. The first effect is due to the laser, which, even if detuned, will heat the electron. The second effect is due to the presence of the phonon bath, which cools down the electron towards the optical potential minimum by phonon emission. The laser heating is governed by the electron-photon scattering rate $\Gamma_{se} = \varepsilon\Gamma_t = \varepsilon\tau^{-1}$, which is written as the product of the probability to find the electron in the excited trion state $\varepsilon$, and the radiative decay rate $\Gamma_t$, i.e. the inverse of the radiative trion lifetime $\tau$. Every scattering event heats the electron up by adding an amount of kinetic energy equal to the in-plane recoil energy $\langle E_R \rangle$, where $\langle \ \rangle$ indicates the average over the emission angle. The corresponding heating rate $R_h$ can be expressed as $R_h = \varepsilon\Gamma_t \langle E_R \rangle$. The laser heating rate does not depend on the lattice temperature, nor on the electron energy. In the figure, the heating rate $R_h$ is shown in orange for different values of the parameter $\varepsilon$. The cooling rate due to the emission of acoustic phonons $R_c$ has been calculated by weighting the phonon emission rate (see Fig. S2) with the phonon energy. We show in blue the $R_c$ for CdTe as a function of the electron energy for a 20 nm quantum well and lattice temperature of 300 mK. An effective equilibrium temperature $k_B T^*$ for the electron can then be defined as the electron energy for which the cooling and heating rate exactly balance each other (i.e. $R_h = R_c$). For the saturation parameters shown in the figure, we obtain equilibrium temperatures in the range 0.1 to 0.25 meV. **(b) Quantum jump simulation.** The figure shows a simulation of how the electron energy changes as a result of photon heating and phonon cooling random events. At each time step in the simulation the electron energy can either remain constant or it can change due to a heating or cooling event with a given probability. The probability of a photon heating event is $1 - e^{-\Gamma_{se} t}$ while the probability of phonon

emission is $1-e^{-\Gamma_{1\to 1}(E)t}$, where $t$ indicates the time delay since the last event. Due to these competing events the electron energy is stochastically fluctuating around a equilibrium energy of about 0.2 meV, but it safely stays below the trap depth of 1.6 meV, shown in orange in the figure. The simulation refers to a 20 nm CdTe QW with saturation parameter $\varepsilon = 0.05$, detuning $\Delta_t = -0.95 E_b$, and lattice temperature of 300 mK.

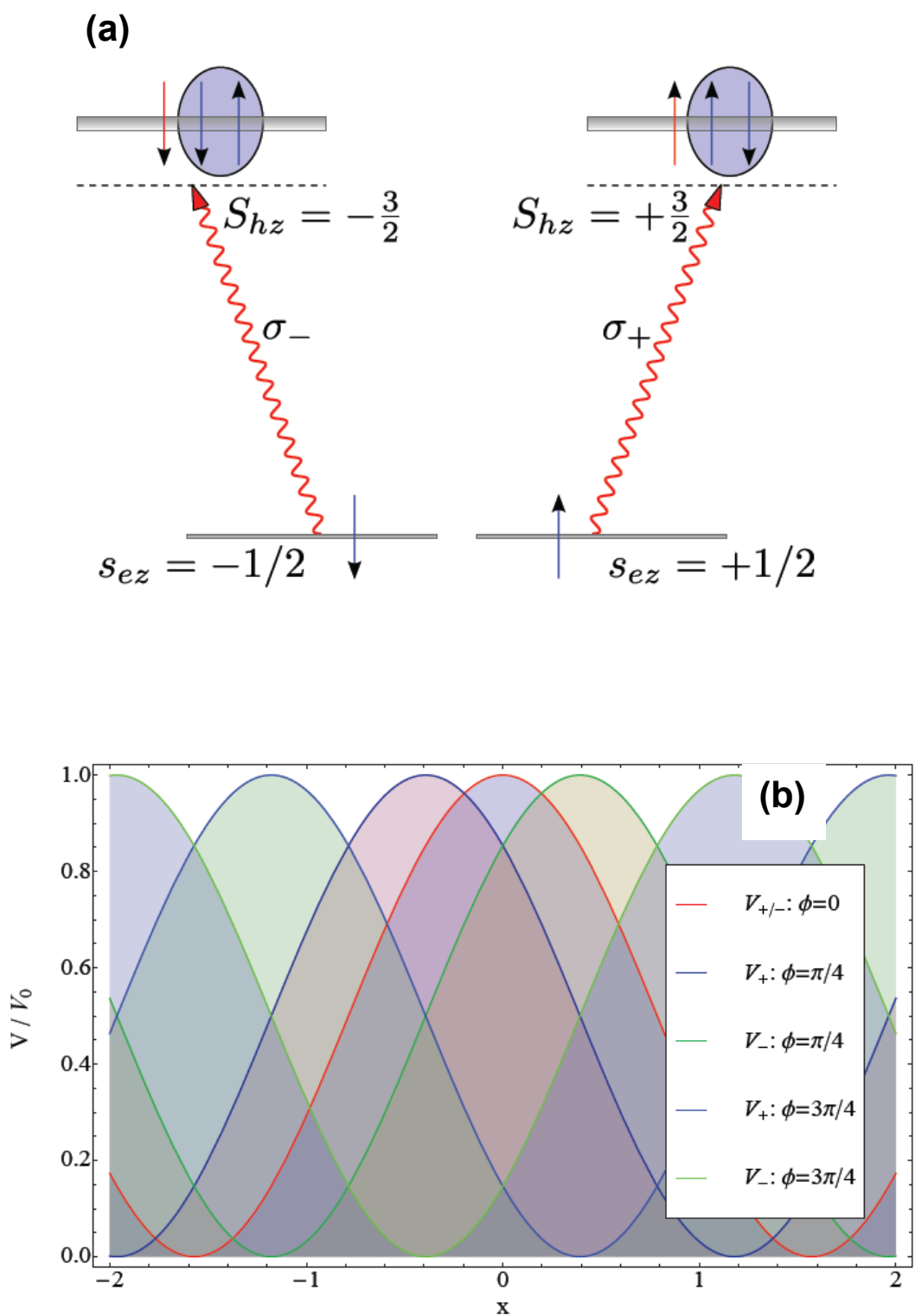

FIGURE S5

FIGURE S5 **(a)** .**Spin effects: Optical selection rules for the electron-trion transition.** Owing to the combined effect of spin-orbit and confinement, the optical potential is spin-selective since the AC Stark shift of the electron level depends on light polarization and on the spin state of the electron. A spin-down electron only couples to $\sigma_-$ polarized photons to form a trion singlet state (left hand side), whereas a spin-up electron is coupled to a trion singlet via $\sigma_+$ polarized photons only (right hand side). The polarization quantization axis is determined by the QW growth direction. Assuming that light hole states are sufficiently separated, which is true for the narrow quantum wells we are considering here, spin-flip Raman transitions are dipole forbidden and the trion system at zero magnetic field appears as a double two-level system. **(b)lin-$\phi$-lin configurations.** The spin-selectivity of the optical potential can be exploited in a so called lin-$\phi$-lin configuration [9,10] realized using two laser beams counter-propagating in the quantum well plane with linear polarization at a relative angle $\phi$. This setup results in a standing wave which can be decomposed into a superposition of a $\sigma_+$- and a $\sigma_-$ polarized standing waves and therefore generates two separate lattice potentials $V_\pm(x,\phi) = V_0 \cos^2(Qx \pm \phi/2)$. By changing $\phi$, one can control the relative separation $\Delta x$ between the two potentials according to $\Delta x = (\phi/\pi)\lambda/2$. The figure shows how the two potentials can be shifted apart by increasing $\phi$. For $\phi = 0$ the two potentials overlap, but as $\phi$ is increased, the $V_+$ potential moves to the left, while the $V_-$ potential moves to the right. Using this approach, electrons could be moved coherently across several lattice sites, as demonstrated for atoms [11,12].